\documentclass[twocolumn,prl,showpacs,preprintnumbers,superscriptaddress,amsmath,amssymb]{revtex4}

\usepackage{graphicx}
\usepackage{dcolumn}
\usepackage{bm}
\begin{document}
%
%
\title{Dephasing Time in a Two-Dimensional Electron Fermi Liquid}
\author{M. Eshkol}
\email{shkolm@post.tau.ac.il}

%
\author{E. Eisenberg}
%
\author{M. Karpovski}
%
\author{A. Palevski}

\affiliation {School of Physics and Astronomy, Raymond and Beverly
Sackler Faculty of Exact Science, Tel-Aviv University, 69978
Tel-Aviv, Israel.}

\date{\today}
\begin{abstract}
The observation of coherent quantum transport phenomena in metals
and semiconductors is limited by the eventual loss of phase
coherence of the conducting electrons on the time scale
$\tau_\varphi$.
We use the weak localization effect to measure the
low-temperature dephasing time
in a two-dimensional electron Fermi liquid in GaAs/AlGaAs
heterostructures. We use a novel temperature
calibration method based on the integer quantum Hall effect in
order to directly measure the electrons' temperature. The data are
in excellent agreement with recent theoretical results, including
contributions from the triplet channel, for a broad temperature
range. We see no evidence for saturation of the dephasing time
down to $\sim100mK$. Moreover, the zero-temperature dephasing time
is extrapolated to be higher than 4ns.

\end{abstract}
\pacs{73.20.Fz, 03.65.Yz, 73.43.Qt, 73.43.Fj}
\maketitle

The electron dephasing time, $\tau_\varphi$, is a quantity of great
importance for the analysis of transport in semiconductor and metal
mesoscopic samples. Essentially,
$\tau_{\varphi}$ sets the scale at which the quantum-mechanical
properties of the microscopic system crossover to the familiar
classical behavior seen in macroscopic objects. The study of
quantum coherence has attracted much attention, motivated both by
questions of fundamental scientific interest concerning sources of
decoherence in materials, and by the possibility of using
solid-state electronic devices to store quantum information. The
investigation of electron dephasing has advanced significantly
thanks to the observation of a variety of prominent quantum
interference phenomena. Weak localization
\cite{Anderson1,Bergmann}, universal conductance fluctuations
\cite{Altshu1,LeeStone}, the Aharonov-Bohm effect \cite {AB} and
persistent currents \cite{Persistent} exhibited in mesoscopic
electronic systems make these systems suitable for studying
decoherence. The most prominent interference effect is
weak-localization, the quantum-mechanical enhancement of coherent
backscattering. This coherent interference is destroyed by the
break of time-reversal symmetry, resulting in a noticeable
"anomalous" magnetoresistance of disordered conductors at low
temperatures and low magnetic fields. Analysis of the
magnetoresistance curves may provide quantitative information on the
various electron dephasing mechanisms.
\par
A number of basic microscopic dephasing processes may coexist in real systems
at low temperatures,
with one or two mechanisms typically dominating, depending on
system dimensionality, level of disorder and temperature. For
two-dimensional semiconducting samples at low temperatures, the
dominating dephasing process is quasi-elastic e-e interactions.
These give rise to $1/\tau_{\varphi}\simeq T^{2}\ln(T)$ at
relatively high temperatures, due to large energy transfer
processes (or, using the terminology of Ref. \cite{Aleiner},
the ballistic term) and
$1/\tau_{\varphi}\simeq T$ at lower temperatures, where small
energy transfer processes dominate the dephasing (diffusive term
in \cite{Aleiner}). Accordingly, the zero temperature dephasing time,
$\tau_\varphi^{0}\equiv\tau_\varphi(T\rightarrow0)$,
is expected to diverge. Contrary to
this prediction, a number of experimental groups have shown
indications for a
finite saturated dephasing time at low temperatures \cite
{Saturation}. Recently, this contradiction has been the focus of
considerable attention. Among the current opinions on the matter,
it has been suggested that the saturated value should depend on
the specific sample geometry \cite {geometry}, the level of
disorder in the sample \cite {disorder}, the microscopic qualities
of the defects \cite {Imry,2CK}, or e-e scattering mediated by the
magnetic exchange interaction \cite{exchange}. Others argue that
the saturation is caused by \emph{extrinsic} mechanisms, such as
magnetic spin-spin scattering \cite{magnetic}, hot electron
effects \cite{Hot e}, electromagnetic noise sources \cite{Altshu1}
or non-equilibrium effects \cite{Ovadiahu}. The possible extrinsic
mechanisms urge caution when determining the actual temperature of
the two-dimensional electron system and ensuring outside radiation
is small.

Most of the above-mentioned experiments were compared with theoretical
results for the two-dimensional electron gas, focusing on the universal
contribution of the singlet channel interaction, both in
the energetically diffusive \cite{Altshu2,Fukayama} and ballistic
regimes \cite{Fukayama,DeSarma}. Recently, the effect of Fermi
liquid renormalization of the triplet channel of the Coulomb
interaction on the dephasing time has been studied theoretically
for arbitrary relation between inverse temperature and elastic
mean free time \cite{Aleiner}. The prefactors of these dependencies
are not universal, but are determined by the Fermi liquid constant
characterizing the spin-exchange interaction. It is expected that
taking into account the Fermi liquid normalization would facilitate
better quantitative understanding of the experimental data.
\par
In this work, weak-localization magnetoresistance measurements
were performed in two-dimensional Fermi liquid fabricated in
GaAs/Al$_{0.3}$Ga$_{0.7}$As heterostructures with high
conductance, in order to extract the dephasing time at various
temperatures down to $\sim$100mK. We compare our results to the
theoretical prediction that includes contributions from both the
singlet and triplet channels. Our measurements are in the
intermediate temperature range, where both small and large energy
transfer scattering contribute to phase braking. The measurements
were accompanied by integer quantum Hall measurements showing
variable-range-hopping behavior in the diagonal resistivity minima
at very low temperatures. This predicted, exponential behavior was
used to calibrate the electrons' temperature as a means to
quantify hot electrons effects. We observe good quantitative
agreement with theory over all the temperature range, in both
energetically ballistic and diffusive regimes. No indications for
saturation of the dephasing time are detected down to the lowest
temperature measured.

\par
It has been shown in \cite{Aleiner} that at low temperatures, where small
energy transfer scattering processes dominate
$(k_BT\tau/\hbar\ll1)$, the temperature dependence of the
dephasing time is:
\begin{eqnarray}
1/\tau_{\varphi}=
\left\{%
1+\frac{3(F_o^\sigma)^{2}}{(1+F_o^\sigma)(2+F_o^\sigma)}
\right\}%
\frac{k_{B}T}{g\hbar}\ln[g(1+F_o^\sigma)]
\nonumber\\
+
\frac{\pi}{4}\left\{1+\frac{3(F_0^\sigma)^{2}}{(1+F_o^\sigma)^{2}}
\right\}%
\frac{(k_{B}T)^2}{\hbar E_{F}}\ln(E_F\tau/\hbar),
\end{eqnarray}
where $F_o^\sigma$ is the interaction constant in the triplet
channel which depends on interaction strength
\cite{Aleiner2,Landau}, $g\equiv2\pi\hbar/e^2R_\Box$ and $E_F$ is
the Fermi energy. At higher temperatures where large energy
transfer scattering processes contribute $(k_BT\tau/\hbar\gg1)$,
\begin{eqnarray}
1/\tau_\varphi = \frac{\pi}{4}\frac{(k_BT)^2}{BE_F}
 \left\{ \ln\left ( \frac{2E_F}{k_BT}\right ) \right. \nonumber\\
 +\left. \frac{3(F^\sigma_o)^2}{(1+F^\sigma_o)^2}\ln\left(\frac{E_F}{k_B T\sqrt{b(F^\sigma_o)}}\right )%
\right\},
\end{eqnarray}
where $b(x)\approx(1+x^2)/(1+x)^2$, and $B$ is a numerical factor
that varies between 0.84 for weak magnetic fields
$(\Omega_H\tau_\varphi\gg1$ where $\Omega_H=4DeH/\hbar c)$ and
0.79 in the opposite limit \cite{Aleiner}. These results were
recently compared by Minkov {\it et al.} \cite{Germanenko} to
measurements of magnetoresistance and dephasing times for samples
of intermediate conductances, where higher orders in $1/g$
contribute. Taking into account high orders corrections, good
agreement between theory and experiment has been observed.
\par

\begin{figure}
\includegraphics[width=0.7\hsize ] {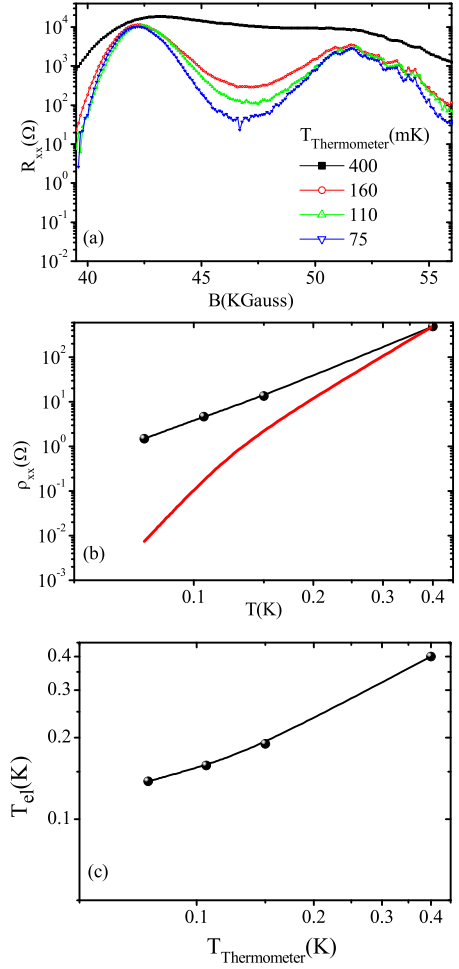}
\caption{\label{fig:f1} (Color online) The temperature calibration
process: (a) Magnetoresitance measurements. The longitudinal
resistance is presented as a function of the magnetic field for
several different temperatures. The magnetic field is in the range
3.9-5T (corresponding to plateau i=3). The temperature ranges from
400mK (top) to around 75mK (bottom), as measured by the
thermometer. (b) The measured resistance minima as function of
thermometer temperature (black circles), compared to the
variable-range-hopping result
$\varrho_{xx}\propto1/T\exp{-(T/T_o)^{1/2}}$ (red solid line).
Clearly, the measured resistance surpasses the
variable-range-hopping results, reflecting the hot electrons
effect. (c) The actual electrons' temperature as function of the
thermometer temperature. }
\end{figure}

The samples are fabricated from single-well AlGaAs/GaAs
heterostructures in order to avoid complications from inter-valley
scattering magnetic impurities, and due to the negligible
spin-orbit coupling in these heterostructures.
The samples are mesa-etched into standard
Hall-bar configuration using standard lithography. The samples
dimensions are $200\mu m$ long and $10\mu m$ wide. The electron
density was $2.8\times10^{11} cm^{-2}$ with a mobility of $87000
 cm^2/Vsec$. The corresponding electron diffusion constant(D) and
mean free time ($\tau$) are $D=0.085m^2/sec$ and
$\tau=3.3\times10^{-12}sec$. The magnetoresistance measurements
are carried out using a four-probe configuration, using a lock-in
amplifier by applying a magnetic field perpendicularly to the
sample. $V_L$, the applied bias on the whole sample of length $L$, is kept
below the temperature \cite{Ovadiahu}, $eV_{L}/k_{B}<T$, rather
than the conventional $eV_{\phi}/k_{B}<T$ criterion, where
$V_{\phi}$ is the bias applied to the phase-coherent length,
$L_{\phi}$, in order to prevent any non-equilibrium effects from
causing dephasing. In addition, we explicitly verified that the
magnetoresistance curve was insensitive to  further reduction in
the voltage bias.
\par
At very low temperatures, lack of good thermal contact between the
lattice and the electrons might occur. This might lead to a
difference between the actual electron temperature and that
measured by the thermometer. This hot electrons effect, requires
careful temperature measurement. We employ longitudinal resistance
measurements in the integer quantum Hall effect regime in order to
directly measure the temperature of the electron gas using an
effect independent of the weak localization phenomenon. It is well
established \cite{Ono} that the longitudinal conductance in the
plateau area in the quantum Hall regime is due to thermal
activation over the mobility edge at relatively high temperatures,
and to variable-range-hopping at lower temperatures. These effects
predict exponentially strong temperature dependence of the
conductivity/resistivity, $\rho_{xx}\propto 1/T\exp
(-(T/T_{0})^{1/2})$. This dependence was measured and shown in
AlGaAs/GaAs heterostructures very similar to ours in Ebert
\emph{et al.} \cite{Klitzing}, at least down to 30mK. We use these
theory and experimental findings to calibrate our temperature by
comparing our data (Fig. 1(a)) with the theoretical prediction
they established. By taking the minima resistivity measured by us
and comparing it to a value from the equation given in Ebert
\emph{et al.}, we measure the electrons' temperature and indeed
find that it is higher than the thermometer temperature,
indicating hot electron effects. In order to minimize small
lock-in amplifier deviations, we calibrated it by setting the
resistivity values at the minima corresponding to plateau i=4 to
zero, where the value is already at the saturated value for the
entire temperature range. In addition, we normalized the measured
and calculated resistivity values at the high temperatures where
we expect the temperature deviation between the gas and
thermometer to be absent in order to fix the prefactors. The
difference between the measured and calculated values is shown in
Fig. 1(b). By comparing the measured data with the theoretical
predictions, we can measure the actual electron gas temperature
(Fig 1(c)).

\begin{figure}
\includegraphics[width=0.90\hsize ]{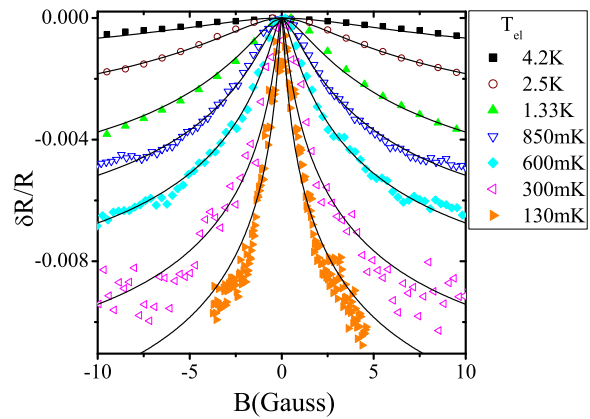}
\caption{\label{fig:f0}(Color online) Weak localization
magnetoresistance measurements at different temperatures. The
temperature range is from 4.2K  (top) down to 130mK(bottom). The
black solid lines are the best fits to Eq. 3.}
\end{figure}

\begin{figure}
\includegraphics[width=0.9\hsize ]{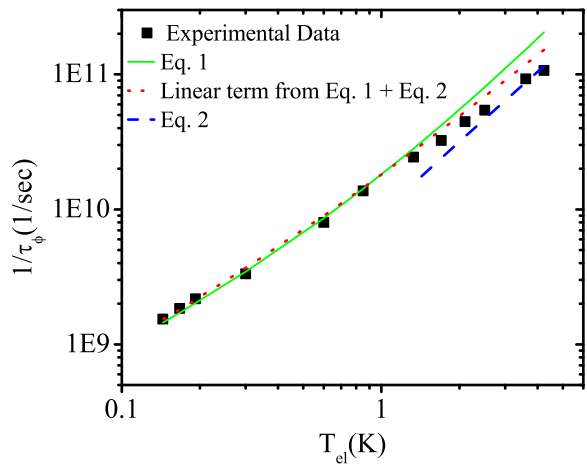}
\caption{\label{fig:f3}(Color online) The temperature dependence
of the dephasing rate $\tau_\phi^{-1}$ extracted from the
weak-localization measurements (Black squares). The green solid
line is the theoretical value from Eq. 1. The blue dashed line is
the theoretical value from Eq 2. The red dotted line is the
theoretical value from the combination of the linear term in Eq. 1
and Eq. 2.}
\end{figure}

According to theory of weak localization, the magnetoconductance
in the 2D limit is given by the following combination of digamma
functions \cite{Hikami}:
\begin{eqnarray}
\Delta\sigma=\frac{e^2}{2\pi^2\hbar}
 \left\{
\frac{3}{2}\Psi \left[ \frac {1}{2}+\frac{B_2}{B} \right] -\Psi
\left[\frac{1}{2}+\frac{B_1}{B} \right] \right.
\nonumber\\
\left.
 -\frac{1}{2} \Psi \left[ \frac{1}{2}+\frac{B_3}{B} \right]
-\ln\left[\frac{B_{2}^{3/2}}{B_1{B_3}^{1/2}}\right] \right\},
\end{eqnarray}
where $\Psi$ is the digamma function and
\begin{eqnarray}
B_1&=&B_0+B_{so}+B_s \nonumber\\
B_2&=&B_\varphi+4/3B_{so}+2/3B_s\nonumber\\
B_3&=&B_\varphi+2B_s. \label{Bs}
\end{eqnarray}
In Eq. (\ref{Bs}), $B_x\equiv\hbar/4eD\tau_x$ are the
characteristic fields of elastic scattering $(B_0)$, spin-orbit
$(B_{so})$, phase loss $(B_\varphi)$ and magnetic impurities
$(B_s)$ related to the corresponding times
$\tau,\tau_{so},\tau_\varphi$ and $\tau_s$. B is the applied
perpendicular field. The Magnetoresistance data are shown in Fig.
2, for temperatures between 4.2K and $\sim$130mK. The solid lines
are best fits using Eq. 3. In our MBE grown samples there are no
magnetic impurities and $B_{so} \ll B_\phi$, making $\tau_\phi$
the only fitting parameter. The values of the extracted dephasing
time from Eq. 3 are plotted in Fig. 3 as a function of
temperature. The green solid line is the theoretical value from
Eq. 1, applicable where the small-energy transfer term dominates,
and the blue dashed line is the theoretical value from Eq. 2,
applicable where the large-energy transfer term dominates. The red
dotted line is the combination of the theoretical value from Eq. 2
and the linear term from Eq. 1, which represents the ballistic
limit with some contribution from the small-energy transfer linear
term. These curves are plotted with no fitting parameter. The
value used for the Fermi-liquid constant is $F_o^\sigma=-0.4$,
consistent with the known value for GaAs for our electron
concentration. This value was used for all theoretical values
described in Fig. 3.

The measured dephasing times agree well with Eq. 1, up to
$T\sim1K$. This is in agreement with the estimated transition
temperature $T=\hbar/k_B\tau\approx1.4K$ describing the transition
to the ballistic limit where large energy transfer processes
dominate. At higher temperatures, comparison to the ballistic term
(Eq. 2) shows agreement at least asymptotically. Combining the
high energy transfer term from the high temperature limit together
with the linear term from Eq. 1, we can observe even better
agreement, albeit with a small deviation at the highest
temperatures which might be the result of the proximity to the
limit where $L_\phi\approx l$, making the application of Eqs. 1,2
somewhat problematic.
\par
The excellent agreement at the low temperature range allows us to
use this equation to extrapolate the dephasing times to lower
temperatures. We estimate our experimental error to be no more
than 10 percent. Using this estimate and attributing the
deviations at the lowest temperature achieved to a constant value,
we can calculate the minimum saturated dephasing time value and
estimate the saturation temperature. The dephasing rate we measure
at the lowest temperature is $1.54ns^{-1}$, while the theoretical
prediction from Eq. 1 is $1.46ns^{-1}$. Taking into account a
possible measurement error of 10\%, and attributing all the
deviation from the predicted theoretical value to an unknown
temperature-independent dephasing mechanism yields a rate of
$0.23ns^{-1}$ for this zero-temperature dephasing mechanism. The
minimal saturated dephasing time is thus estimated to be
$\tau_\varphi^{sat}>4ns$ and the maximal corresponding saturation
temperature is $\sim25mK$. To the best of our knowledge, this
saturated dephasing time value is higher then the saturation
dephasing times reported in previous experiments.

To conclude, we have measured the dephasing time using weak
localization magnetoresistance measurement, demonstrating very
good quantitative agreement with recent theoretical results for a
Fermi liquid (given in Eqs. 1 and 2), with no fitting parameters.
Our data are at a range where both large and small energy transfer
scattering contribute to dephasing. We demonstrate the agreement
on a relatively broad temperature scale. We see no evidence for
saturation down to the lowest temperature measured. Comparing our
data to the theoretical results, we limit the possible temperature
independent dephasing rate to $0.23 ns^{-1}$ at most, resulting in
zero-temperature dephasing time of at least 4ns, which cannot be
observed in our samples at electron temperatures above 25mK.
\par
We would like to thank I. L. Aleiner, K. B. Efetov, A. D. Zaikin
and G. Sch\"{o}n for fruitful discussions. The support of the
Israel Science Foundation founded by the Israel Academy of
Sciences and Humanities, Centers of Excellence Program is
gratefully acknowledged. E.E. is supported by an Alon fellowship at
Tel-Aviv University.

\end{document}